\documentclass[12pt]{article}
\usepackage{graphicx}

\newcommand\pubnumber{Article 16 in
eConf C1304143}
\newcommand\pubdate{\today}

\def\penn{Department of Astronomy \& Astrophysics\\
The Pennsylvania State University, Univerisity Park, PA, USA}
\def\swri{Space Science \& Engineering Division \\
Southwest Research Institute, 6220 Culebra Rd, San Antonio, TX, USA}
\def\Title#1{\begin{center} {\Large #1 } \end{center}}
\def\Author#1{\begin{center}{ \sc #1} \end{center}}
\def\Address#1{\begin{center}{ \it #1} \end{center}}
\def\andauth{\begin{center}{and} \end{center}}

\def\accepted#1{\begin{center}Accepted to {\sl #1} \end{center}}
\newcommand\pubblock{\rightline{\begin{tabular}{l} \pubnumber\\
         \pubdate  \end{tabular}}}
\newenvironment{Abstract}{\begin{quotation}  }{\end{quotation}}
\newenvironment{Presented}{\begin{quotation} \begin{center} 
             PRESENTED AT\end{center}\bigskip 
      \begin{center}\begin{large}}{\end{large}\end{center} \end{quotation}}

\def\beq{\begin{equation}}
\def\eeq#1{\label{#1}\end{equation}}
\def\eeqn{\end{equation}}

\def\beqa{\begin{eqnarray}}
\def\eeqa#1{\label{#1}\end{eqnarray}}
\def\eeqan{\end{eqnarray}}

\let\bar=\overbar

\def\Dslash{\not{\hbox{\kern-4pt $D$}}}
\def\dslash{\not{\hbox{\kern-2pt $\del$}}}

\def\msb{{\bar{\ssstyle M \kern -1pt S}}}

\begin{document}
\begin{titlepage}
\pubblock

\vfill
\accepted{The Astrophysical Journal}
\Title{GRB Flares:  A New Detection Algorithm, Previously Undetected Flares, and Implications on GRB Physics}
\vfill
\Author{ Craig A. Swenson}
\Address{\penn}
\andauth{}
\Author{ Peter W. A. Roming}
\Address{\swri}
\vfill

\begin{Abstract}
Flares in GRB light curves have been observed since shortly after the discovery of the first GRB afterglow. However, it was not until the launch of the Swift satellite that it was realized how common flares are, appearing in nearly 50\% of all X-ray afterglows as observed by the XRT instrument. The majority of these observed X-ray flares are easily distinguishable by eye and have been measured to have up to as much fluence as the original prompt emission. Through studying large numbers of these X-ray flares it has been determined that they likely result from a distinct emission source different than that powering the GRB afterglow. These findings could be confirmed if similar results were found using flares in other energy ranges. However, until now, the UVOT instrument on Swift seemed to have observed far fewer flares in the UV/optical than were seen in the X-ray. This was primarily due to poor sampling and data being spread across multiple filters, but a new optimal co-addition and normalization of the UVOT data has allowed us to search for flares in the UV/optical that have previously gone undetected. Using a flare finding algorithm based on the Bayesian Information Criterion, we have analyzed the light curves in the Second UVOT GRB Catalog and present the finding of at least 118 unique flares detected in 68 GRB afterglows.
\end{Abstract}
\vfill
\begin{Presented}
``Huntsville in Nashville" GRB Symposium\\
Nashville, TN, USA,  April 14--18, 2013
\end{Presented}
\vfill
\end{titlepage}
\def\thefootnote{\fnsymbol{footnote}}
\setcounter{footnote}{0}

\section{Introduction}

Flares have been observed in GRB light curves since shortly after the discovery of the X-ray afterglow\cite{Piro:1997zu}. In the Swift era, nearly 50\% of all afterglows have some flaring in afterglows observed by the XRT. Most of these are large and easily distinguishable by eye, and these large flares have been the subject of several studies\cite{Chincarini:2007fp,Falcone:2007rz}. UV/optical flares are seen less frequently in the UVOT data, primarily due to poor sampling and data being spread across multiple filters. These issues have been addressed by the Second \emph{Swift} Ultraviolet/Optical Telescope GRB Afterglow Catalog\cite{Roming:2013}. We developed a new flare finding algorithm based on the Bayesian Information Criterion to identify previously undetected flares in the UV/ optical and to also identify smaller flares in the X-ray that have been overlooked.

\section{Methodology}

The Second \emph{Swift} Ultraviolet/Optical Telescope GRB Afterglow Catalog makes use of optimal co-addition and normalization to maximize the temporal coverage of GRB light curves.  Optimal co-addition is a process that optimally weights each exposure in order to maximize the signal-to-noise-ratio (S/N), which decreases as the source count rate becomes low and the background relatively high.  The method takes into account the decaying nature of the GRB light curve, predicting the expected count rate over time and calculates the correct amount of image co-addition necessary to produce the maximum S/N.

	However, even with the increased temporal coverage that comes with optimally co-adding and normalizing the data, we still require a statistically robust method to confirm the existence of the much smaller flares, as compared with the X-ray, that we hope to detect in the UV/optical.  For this study we have used the publicly available \texttt{R}\cite{R} package \texttt{strucchange} \cite{Zeileis:2002} and the breakpoints analysis function contained within the package \cite{Zeileis:2003}.  This method employs the use of the Bayesian Information Criterion (BIC) \cite{Schwarz:1978} to determine the presence of \emph{breakpoints} in time series data that indicate the need for additional fitted components to the data set; in our case we interpret those identified components as belonging to flares.

Generally, the Bayesian Information Criterion (BIC) is a ``summary of the evidence provided by the data in favor of one scientific theory...as opposed to another''\cite{Kass:1995}. Given a number of possible different fits to a data set, the BIC can be used to determine the best fit to the data. We can detect the presence of flares in the GRB afterglow by iteratively fitting the light curve and calculating the BIC for each fit using the breakpoints analysis function. Overfitting the data results in a penalized BIC, ensuring the optimal fit with the fewest additional parameters. For each GRB afterglow we perform 10,000 Monte Carlo simulations, each time identifying potential flares and determining the optimal fit. Potential flares are assigned a measure of confidence based on the number of times they are detected during the 10,000 simulations.

\begin{figure}[htb]
\centering
\includegraphics[width=\textwidth]{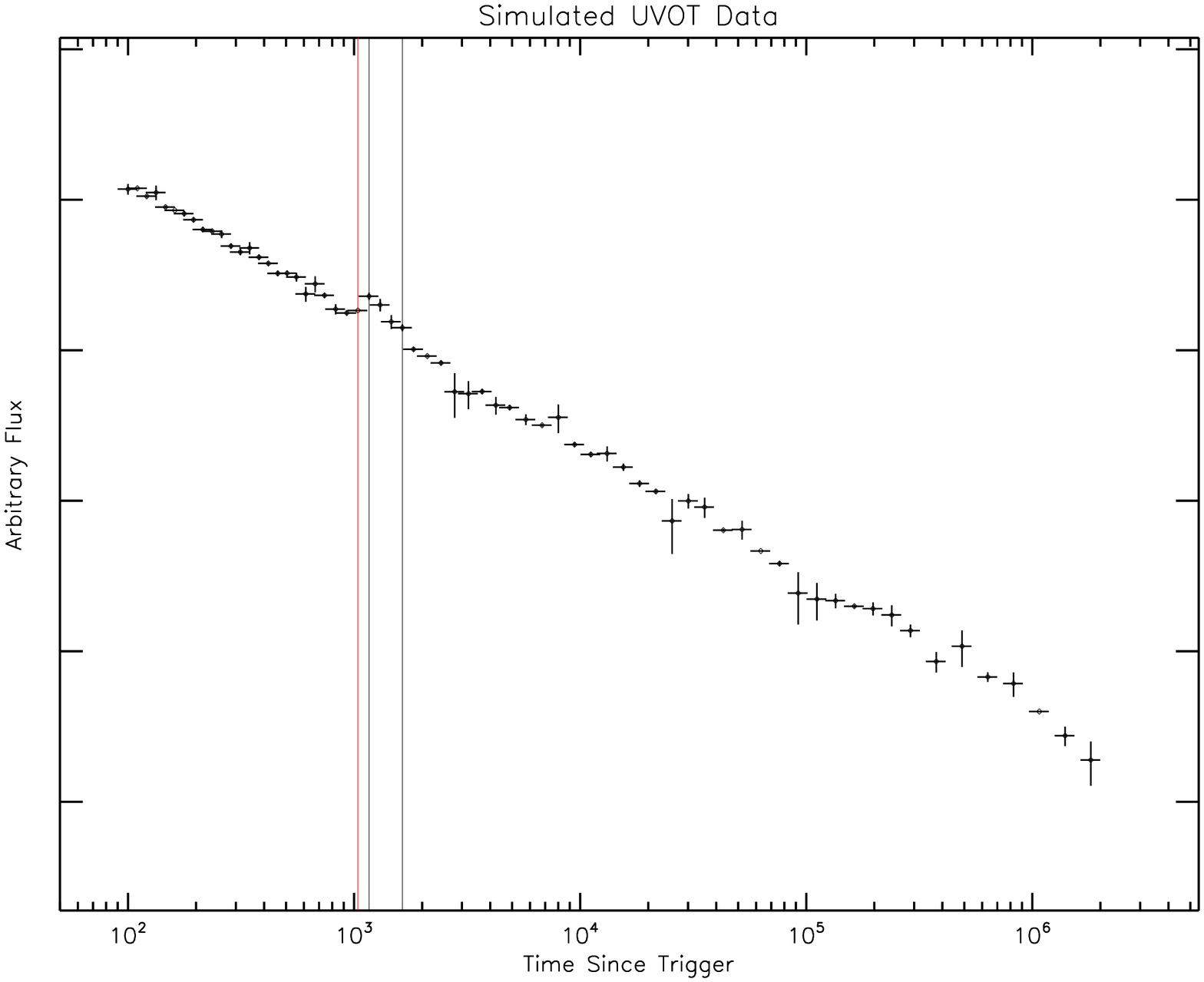}
\caption{Simulated UVOT data showing a flare with a small amplitude and duration.  Due to the abrupt rise to the peak of the flare, the code does not identify a unique point as being associated with the start of the flare.  We assign the first point prior to $T_{peak}$ to be $T_{start}$.  The data point assigned as $T_{start}$ is identified by the red line.}
\label{fig:sim_data}
\end{figure}

Figure~\ref{fig:sim_data} shows an example light curve, using simulated UVOT data, to show how the algorithm detects a single flare.  Each of the black vertical lines represents a point in the light curve where the algorithm detected a deviation that required an additional component be added to the overall fit to the light curve.  These two points are the peak of the flare and the end of the flare where it rejoins the underlying light curve.  Because the rise time to peak flux of the flare is so short, the algorithm doesn't automatically detect the beginning of the flare and our flare finding code therefore automatically assigns the flare start time (red vertical line).

\begin{figure}[htb]
\centering
\includegraphics[width=0.8\textwidth,angle=90]{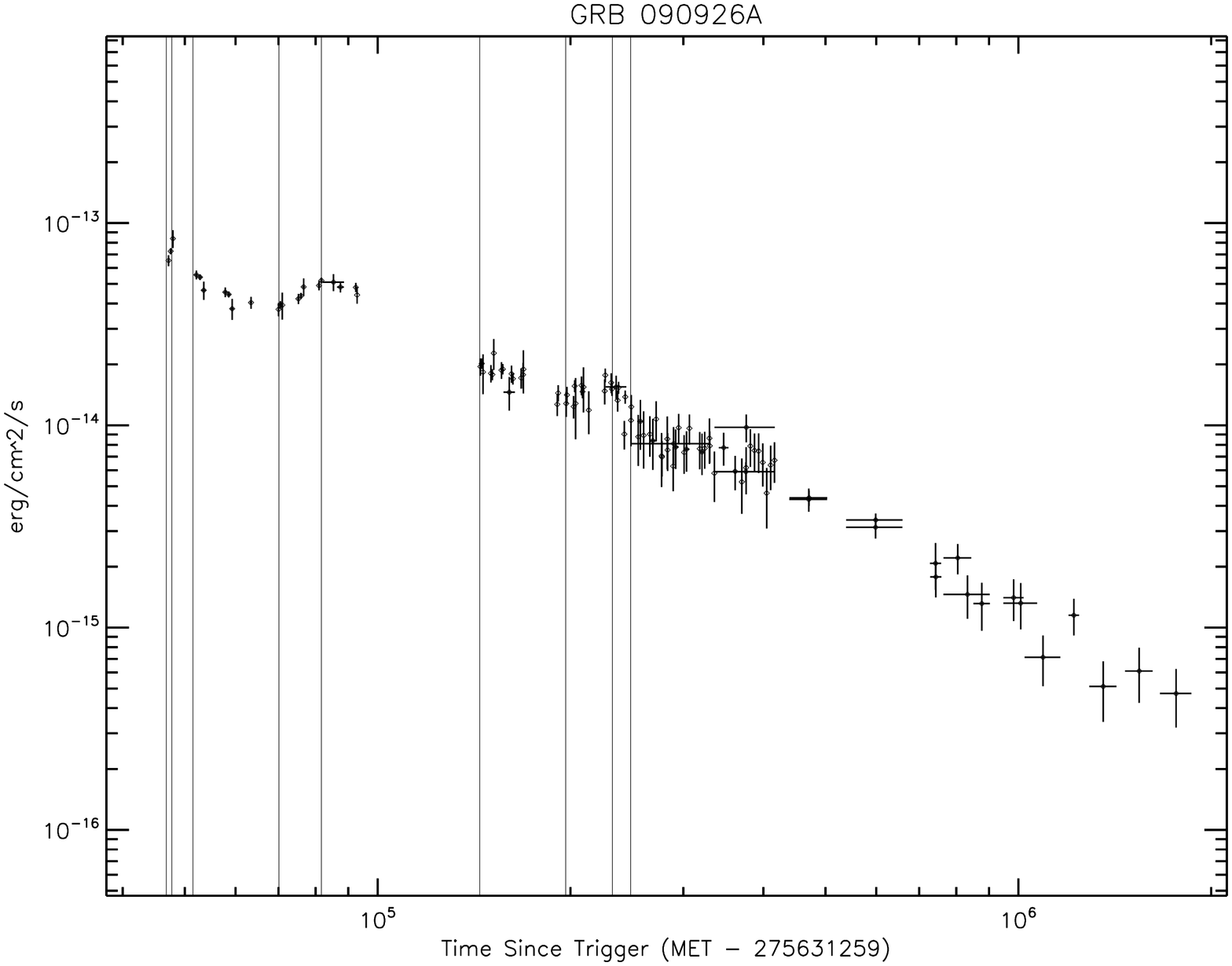}
\caption{Optimally co-added and normalized UVOT light curve of GRB 090926A showing the 9 breakpoints detected by the flare finding algorithm.  The 9 breakpoints are combined to form 3 individual flares.  Overlapping time bins are due to the normalization of the original multi-filter observations.}
\label{fig:uvot_lightcurve}
\end{figure}

Figure~\ref{fig:uvot_lightcurve} shows the results of running the flare finding algorithm on the co-added and normalized light curve of GRB 090926A, a GRB with multiple late-time flares.  This example also shows the difficulties caused by observing constraints (the large gaps in the data) and how the method deals with those difficulties (by assigning the last/first data point before/after the gap).

%%%%%%%%%%%%%%%%%%%%%%%%%%%%%%%%%%%%%%%%%%%%%%%%%%%%%%%%%%%%%%%%%%%%%%%%%
%%
%%   use this format to include an .eps figure into your paper
%%
%\begin{figure}[htb]
%\centering
%\includegraphics[height=1.5in]{magnet}
%\caption{Plan of the magnet used in the mesmeric studies.}
%\label{fig:magnet}
%\end{figure}
%%%%%%%%%%%%%%%%%%%%%%%%%%%%%%%%%%%%%%%%%%%%%%%%%%%%%%%%%%%%%%%%%%%%%%%%%%%

\section{Results and Conclusions}

We analyzed 201 UVOT GRB light curves and found potential flaring episodes in 68 of the light curves. We found an average of ~2 flares per GRB for those that exhibited flaring. The flaring is generally restricted to the first 1000 seconds of the afterglow, but can be observed and detected to beyond $10^5$ seconds. 15\% of the flares peaked more than 1000 seconds after the trigger and the majority of these were also among the strong flares with flux ratios $>$ 2. This may be an observational bias, but could also indicate the presence of some sort of energy build-up that causes large flares at late times or even an alternate emission mechanism from those flares observed early in the light curve. More than 80\% of the flares detected have a $\Delta$t/t $<$ 0.5. Flares were observed with flux ratios relative to the underlying light curve of between 0.04 and 55.42.

We have also analyzed 548 XRT light curves in order to identify smaller X-ray flares that have been previously overlooked.  Our initial findings show large numbers of potential flaring periods, many of them small, in addition to the previously identified and studied large flares.  We are now verifying the potential flaring periods, calculating the flux ratios and $\Delta$t/t values in preparation for publishing a GRB X-ray flare catalog.  Once complete we will perform a complete cross-correlation analysis examining the relationship between the UV/optical and X-ray flaring.

Many studies have shown that the source of GRB flares is likely linked to the internal engine.  Currently the precise nature of the GRB internal engine remains largely unknown.  The complementary nature of these two flare catalogs will allow for more stringent constraints on the origin of flares in GRBs through cross-correlation of these two energy regimes.  Understanding GRB flares is crucial to our unlocking the mystery of the GRB central engine.

%%%%%%%%%%%%%%%%%%%%%%%%%%%%%%%%%%%%%%%%%%%%%%%%%%%%%%%%%%%%%%%%%%%%%%%%%
%%
%%   use this format to include a LaTeX table  into your paper
%%
%\begin{table}[t]
%\begin{center}
%\begin{tabular}{l|ccc}  
%Patient &  Initial level($\mu$g/cc) &  w. Magnet &  
%w. Magnet and Sound \\ \hline
% Guglielmo B.  &   0.12     &     0.10      &     0.001  \\
% Ferrando di N. &  0.15     &     0.11      &  $< 0.0005$ \\ \hline
%\end{tabular}
%\caption{Blood cyanide levels for the two patients.}
%\label{tab:blood}
%\end{center}
%\end{table}
%%%%%%%%%%%%%%%%%%%%%%%%%%%%%%%%%%%%%%%%%%%%%%%%%%%%%%%%%%%%%%%%%%%%%%%%%%%

\end{document}